\documentstyle[aps,epsfig]{revtex}

\def\be{\begin{equation}}

\def\ee{\end{equation}}

\def\lsim{\raise0.3ex\hbox{$<$\kern-0.75em\raise-1.1ex\hbox{$\sim$}}}

\def\gsim{\raise0.3ex\hbox{$>$\kern-0.75em\raise-1.1ex\hbox{$\sim$}}}

\def\NP{{ Nucl.\ Phys.\ }}

\def\PL{{ Phys.\ Lett.\ }}

\def\PRL{{ Phys.\ Rev.\ Lett.\ }}

\begin{document}

\vskip 1 cm

\centerline{\large{\bf Site Percolation and Phase Transitions in Two Dimensions}}

\vskip 1.5cm

\centerline{\bf Santo Fortunato}

\bigskip

\centerline{Fakult\"at f\"ur Physik, Universit\"at Bielefeld}

\par

\centerline{D-33501 Bielefeld, Germany}

\vskip 1.0cm

\noindent

\centerline{\bf Abstract:}

\medskip

The properties of the pure-site clusters of spin models, i.e. 
the clusters which are obtained by joining nearest-neighbour 
spins of the same sign, are here investigated. In the 
Ising model in two dimensions it is known that such clusters undergo a percolation
transition exactly at the critical point.
We show that this result is valid for a wide
class of bidimensional systems
undergoing a continuous magnetization transition. 
We provide numerical
evidence for discrete as well as for continuous spin models, 
including $SU(N)$ lattice gauge theories. The critical 
percolation exponents do not coincide with the ones of the thermal
transition, 
but they are the same for models belonging to the same
universality class.

\bigskip

\vskip1.5cm

The idea to explain the mechanism of a  
phase transition in terms
of the interplay of geometrical structures
which can be identified in the system is quite old
\cite{onsager}: the growth of such structures close to the
transition represents the increase of the
range of the correlation between different parts of the system,
so that the formation
of a spanning structure corresponds to a divergent correlation
length and to the transition to a new state of
global order for the system.

Here we exclusively 
refer to lattice models and the geometrical structures of interest
are usually {\it clusters} of neighbouring particles (spins).
Percolation theory \cite{stauff} is the ideal framework to deal with 
clusters. 
Near the critical point of the percolation
transition there is power-law behaviour for the percolation variables
and one can define an analogous set of critical exponents as 
for standard thermal transitions. The main percolation 
variables are:

\begin{itemize}
\item{the percolation strength $P$, i.e. the probability that a site chosen
at random belongs to a percolating cluster;}
\item{the average cluster size $S$,
\begin{equation}\label{defS}
  S\,=\,\frac{\sum_{s} {{n_{s}s^2}}}{\sum_{s}{n_{s}s}}~,
\end{equation}
where $n_s$ is the number of clusters with $s$ sites and the sums exclude eventual
percolating clusters.
}
\end{itemize}

One can in principle define the clusters arbitrarily, but, in order
to reproduce the critical behavior of the model, the following
conditions must be satisfied:

\begin{itemize}
\item{the percolation point must coincide with the thermal critical point;}
\item{the connectedness length (average cluster radius) diverges as the 
thermal correlation length (same exponent);}
\item{the percolation strength $P$ near the threshold
varies like the order parameter $m$
of the model (same exponent);}
\item{the average cluster size $S$ diverges as the 
physical susceptibility $\chi$ (same exponent).}
\end{itemize}

The first system
to be investigated was of course the Ising
model. If one takes an Ising configuration 
one can immediately isolate the "classical" magnetic domains, i.e. the clusters formed by
binding to each other
all nearest-neighbouring sites carrying spins of the same sign. 
It turns out that these clusters have indeed interesting properties
in two dimensions, since their percolation temperature
coincides with the critical temperature of the magnetization transition
\cite{connap}; the percolation exponents, however, do not coincide
with the Ising exponents \cite{sykes}.

In this paper we show that the result is quite general in two dimensions,
being valid for several models
undergoing a continuous magnetization transition. 
In particular we shall see that the presence of 
antiferromagnetic interactions does not affect 
the result, at least to the extent of the cases studied here. 
Moreover the critical percolation exponents
do not randomly vary from one model to the other but are uniquely
fixed by the universality class the model belongs to.

Our results are based on Monte Carlo simulations on square 
lattices of various systems
near criticality. 
The models we studied can be divided in two groups:
models with $Z(2)$ global symmetry and a magnetization transition
with Ising exponents and models with $Z(3)$ global symmetry  
and a magnetization transition with exponents belonging to the
2-dimensional 3-state Potts model universality class. The systems belonging
to the first group are:

\begin{enumerate}
\item{the Ising model, ${\cal{H}}=-J\sum_{ij}s_is_j$ \,\,\,($J>0$, $s_i={\pm}1$);}
\item{a model with nearest-neighbour (NN) ferromagnetic coupling and 
a weaker next-to-nearest (NTN) antiferromagnetic coupling: 
${\cal{H}}=-J_1\sum_{NN}s_is_j-J_2\sum_{NTN}s_is_j$ ($J_1>0$,
$J_2<0$, $|J_2/J_1|=1/10$, $s_i={\pm}1$);}
\item{the continuous Ising model, ${\cal{H}}=-J\sum_{ij}S_iS_j$ 
\,\,\,($J>0$, $-1{\leq}S_i{\leq}+1$);}
\item{SU(2) pure gauge theory in 2+1 dimensions.}
\end{enumerate}

The models belonging to the second group are:

\begin{enumerate}
\item{the 3-state Potts model, ${\cal{H}}=-J\sum_{ij}\delta(s_i,s_j)$ \,\,\,($J>0$, $s_i=1,2,3$);}
\item{a model obtained by adding to 1) 
a weaker next-to-nearest (NTN) antiferromagnetic coupling: 
${\cal{H}}=-J_1\sum_{NN}\delta(s_i,s_j)-J_2\sum_{NTN}\delta(s_i,s_j)$ \,\,\,($J_1>0$,
$J_2<0$, $|J_2/J_1|=1/10$, $s_i=1,2,3$);}
\end{enumerate}

In each case we produced the thermal equilibrium configurations by using
standard Monte Carlo algorithms, like Metropolis
or heat bath; whenever we could (Ex. models 1 and 3 of 
the first group, model 1 of the second) we adopted
cluster algorithms in order to reduce the autocorrelation 
time\footnote{For the models 2 of both groups a cluster dynamics  
exists, but it does not bring great advantages because of the 
huge size of the clusters to be flipped at the critical point}.
At each iteration we calculated the energy and the magnetization;
the latter is necessary to study the thermal transition.
For the identification of the pure-site clusters in each configuration
we made use of the algorithm 
devised by Hoshen and Kopelman \cite{kopelman}, with free boundary conditions. 
Finally we determined
the percolation strength $P$, the average cluster size $S$ and the size $S_{M}$
of the largest cluster of each configuration, from which we can calculate
the fractal dimension $D$ of the percolating cluster.
We say that a cluster percolates if it spans the lattice from top 
to bottom.
The finite size scaling laws at the critical temperature $T_p$ for the percolation variables
read 
\begin{eqnarray}
P(T_p)\,&\propto&\,L^{-\beta_p/\nu_p}\\
S(T_p)\,&\propto&\,L^{\gamma_p/\nu_p}\\
S_M(T_p)\,&\propto&\,L^{D},
\end{eqnarray}
where $L$ is the lattice side and $\beta_p$, $\gamma_p$, $\nu_p$ are 
critical exponents of the percolation transition.
Moreover, 
to study the percolation transition it is helpful to define also the
percolation cumulant. It is the probability of having percolation at
a given temperature and lattice size, i.e., the quantity obtained by
dividing the number of 
"percolating" samples by the total number of analyzed configurations. 
This variable has three remarkable properties:

\begin{itemize}
\item{if one plots it as a function of $T$, all curves corresponding to
different lattice sizes cross at the same temperature $T_p$, which marks the
threshold of the percolation transition;}
\item{the percolation cumulants for different values of the lattice size
$L$ coincide, if considered as functions of
$t_pL^{1/\nu_p}$ ($t_p=(T-T_p)/T_p$, $\nu_p$ is the exponent of the percolation
correlation length);}
\item{the value of the percolation cumulant at $T_p$ is a universal quantity,
    i.e. it labels a well defined set of critical indices}.
\end{itemize}

These features allow a rather precise determination of the critical
point and a fairly good estimate of the critical exponent $\nu_p$.
\begin{figure}[htb]
  \begin{center}
    \epsfig{file=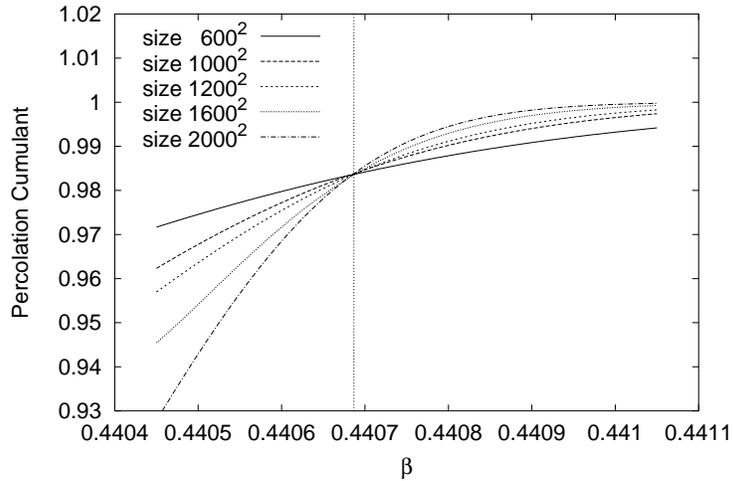,width=10cm}
    \caption{\label{cum}{Percolation cumulant of pure-site 
        clusters in the 2D Ising model as a function of $\beta=J/kT$ for 
        different lattice sizes. The curves cross remarkably at the same point, in 
        agreement with the critical point of the magnetization transition (vertical
        dotted line in the plot).}}
  \end{center}
\end{figure}
We start to expose our results from the models with an Ising-like
transition. We included the Ising model itself in order to reproduce 
the results of \cite{connap,sykes} and to precisely determine 
the percolation exponents. 
We simulated the model on several lattices, from $600^2$ to $2000^2$.
The number of measurements we have taken ranged from a minimum of
20000 for the large $2000^2$ lattice to over 100000 for the others.
From the crossing
of the cumulants (Fig. \ref{cum}) we obtained for the percolation temperature
$J/kT_p=0.44069(1)$, in excellent agreement with the
critical temperature $J/kT_c=\log(1+\sqrt{2})/2=0.44068679...$.
For all models we investigated
it is possible to perform a precise interpolation
of the raw data by applying  
the density of state method \cite{DSM} both to the thermal and to the
percolation variables. The curves in Fig. \ref{cum} are examples of such
interpolations; we did not put the errors to show more clearly the crossing
point and the fact that the expected behaviour is already very well 
represented by the mean values, which is due to the 
high statistics we could reach in this simple case.  

\begin{figure}[htb]
  \begin{center}
    \epsfig{file=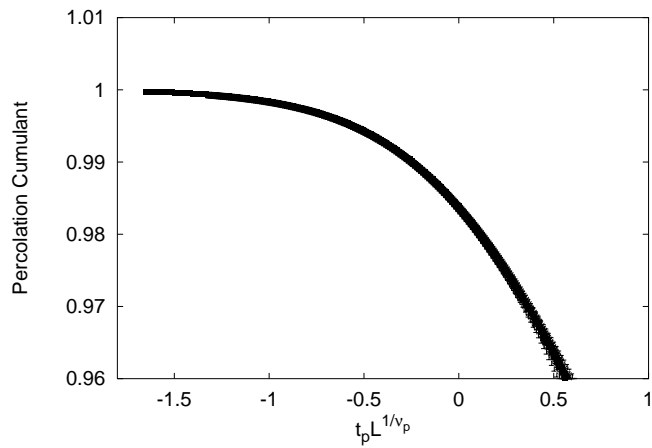,width=10cm}
    \caption{\label{Scalcum}{Rescaling of the percolation cumulant curves of
        Fig. \ref{cum}, by setting $\nu_p=\nu_{Is}=1$.
        The curves fall on top of each other.}}
  \end{center}
\end{figure}

By rescaling the percolation cumulant curves as described above, we obtain
Fig. \ref{Scalcum} if we choose for the exponent $\nu_p$ the value of the 
thermal 2D Ising exponent $\nu_{Is}=1$. This time
we have also drawn the errors, and we see that all data relative to different
lattice sizes fall onto one and the same curve, which shows that
$\nu_p=\nu_{Is}=1$, as
suggested in \cite{CK}.
The percolation exponents and the threshold
value of the cumulant 
are listed in Table \ref{tab1} (Model 1), where we also put the critical percolation indices 
of the other three models of this group. 
The values of $\gamma_p$, $\beta_p$ and of the fractal dimension $D$
are in good accord 
with the theoretical predictions of \cite{stella} (first line of Table \ref{tab1}).

\begin{figure}[htb]
  \begin{center}
    \epsfig{file=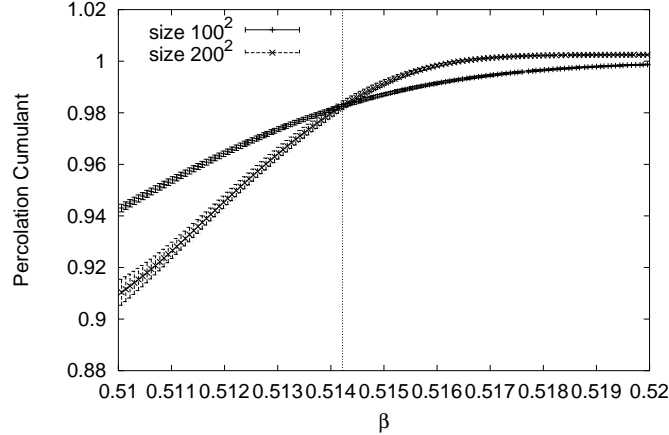,width=10cm}
    \caption{\label{FAcum}{Percolation cumulant of pure-site clusters for the 
$Z(2)$ model with nearest-neighbour ferromagnetic and next-to-nearest-neighbour
antiferromagnetic
interactions.}}
  \end{center}
\end{figure}

For the model with competitive interactions we plot  
the percolation cumulant for two lattice sizes,
$100^2$ and $200^2$ (Fig. \ref{FAcum}). Our statistics
for this system is of 20000 measurements for each temperature and lattice size,
we could not increase it because the algorithms one can use are not
very efficient in this case. However, the result is clear: 
we see that the percolation temperature agrees with the
magnetization one (dotted vertical line in the plot).
More precisely we find $J_1/kT_p=0.51422(12)$, to be
compared with
$J_1/kT_c=0.51418(10)$.  
The values of 
the critical percolation exponents agree with the ones of the 2D Ising model
as we can see in Table \ref{tab1} (Model 2).

\begin{figure}[htb]
  \begin{center}
    \epsfig{file=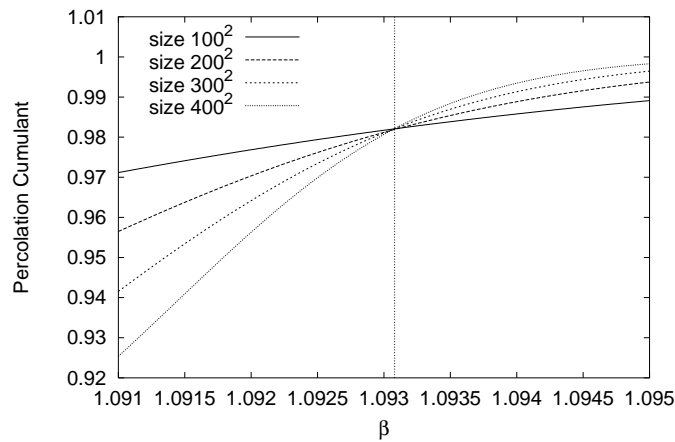,width=10cm}
    \caption{\label{Cumcontis}{Percolation cumulant of pure-site clusters for the 
continuous Ising model.}}
  \end{center}
\end{figure}

The magnetization transition of the continuous Ising model was studied in \cite{santo}.
The value of the critical temperature is $J/kT_c=1.09312(1)$.
For our simulations we made use of the same algorithm described in \cite{santo}, which consists in 
a combination of heat bath steps and cluster flippings; this algorithm is 
quite efficient and we could push our statistics up to 100000 measurements for
each lattice size.
We remark that in this case we have
to do with a continuous spin variable, so that we 
join nearest-neighbouring like-signed spins, although their
absolute values are, in general, different.
Fig. \ref{Cumcontis} shows the percolation cumulant
for the four lattice sizes we took: $100^2$, $200^2$, $300^2$, $400^2$. Due to
our high statistics, we did not draw the errors in our plot, and we see that
the curves of the mean values cross remarkably at the same point, which
coincides with the thermal critical point (dotted vertical line in the figure).
Our estimate
of the percolation temperature is $J/kT_p=1.09311(3)$.

We notice once again that the critical indices
are, within errors, the same we found for the previous two models (Model 3
in Table \ref{tab1}).

\begin{figure}[htb]
  \begin{center}
    \epsfig{file=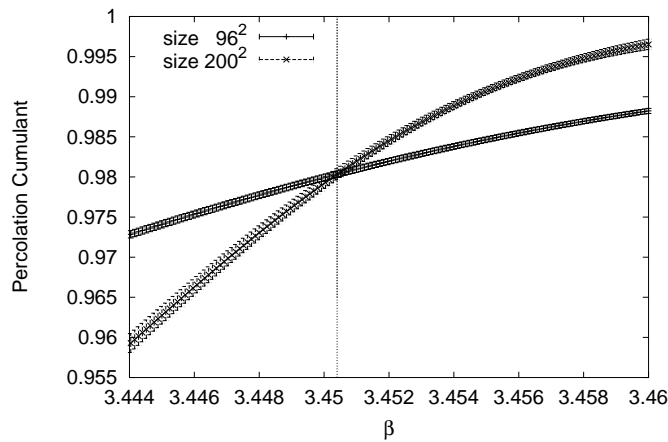,width=10cm}
    \caption{\label{SU2c}{Percolation cumulant of pure-site clusters for 2+1 $SU(2)$
gauge theory, $N_{\tau}=2$.}}
  \end{center}
\end{figure}

Finally we considered the more involved case of $SU(2)$ lattice gauge theory,
in the pure gluonic sector. It is known that in this case there is a 
confinement-deconfinement transition with Ising exponents. The order parameter
is the lattice average $L$ of the 
Polyakov loop, which is a continuous variable like the spins
of the continuous Ising model we examined above. We analyzed 
the 2+1 dimensional case, which corresponds to two space dimensions. The number
of lattice spacings in the imaginary time direction is $N_{\tau}=2$. 
We took four lattices: $64^2$, $96^2$, $128^2$ and $200^2$. 
The statistics ranged from 20000 to 50000 measurements.
Our value for the
critical coupling is $\beta_c=4/{g_c}^2=3.4504(11)$, which is by an order
of magnitude more precise than the value reported in \cite{Teper}. The original 
$SU(2)$ matrix configurations on the links
of the 2+1 dimensional lattice are 
projected onto Polyakov loop configurations on the sites of
a 2-dimensional lattice. For the 
percolation analysis we investigated the Polyakov loop configurations.
In order to build the clusters we proceeded as for the continuous Ising
model, by joining nearest-neighbouring sites carrying like-signed
values of the Polyakov loop variable. 
In Fig. \ref{SU2c} we plotted the percolation cumulant curves as a function
of the coupling $\beta$ for
two lattice sizes, $96^2$ and $200^2$. Also here the coincidence
of the percolation with the critical point seems to be clear.
The percolation coupling we determined is
$\beta_p=4/{g_p}^2=3.4501(14)$.
The critical percolation indices are 
again the same we found so far (Model 4 in Table \ref{tab1}).

We complete our investigation by studying the models of the second group.
Now we have three spin states and the clusters are simply obtained
by binding nearest-neighbouring sites in the same spin state. 
The fact that the pure-site clusters percolate at $T_c$ for the 3-state Potts
model in two dimensions was found, though not rigorously, in \cite{conpe}.
The percolation temperature that we determined is $J/kT_p=1.00511(9)$, in 
agreement with the thermal threshold $J/kT_c=\log(1+\sqrt{3})=1.00505...$.
The exponents of the percolation transition are listed in Table \ref{tab2}
(Model 1).
We see that they are different from the thermal exponents ($\beta/\nu=2/15$,
$\gamma/\nu=26/15$, $\nu=5/6$), except $\nu_p$, which, like in the Ising
case, coincides with the thermal value $5/6$. We remark that the 
values of the critical indices are in very good accord with 
the predictions of \cite{vander} (first line of Table \ref{tab2})).
The percolation exponents are also different from the ones we have found for the
models in the 2D Ising universality class.

\begin{figure}[htb]
  \begin{center}
    \epsfig{file=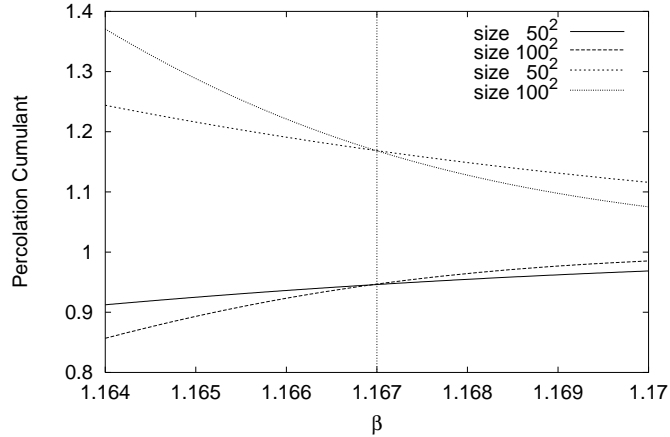,width=10cm}
    \caption{\label{Pottsfr}{Comparison of the percolation point of pure-site
clusters with the magnetization point for the $Z(3)$ model with competitive
interactions. The upper curves represent the Binder cumulant and the lower ones
the corresponding percolation cumulant. The temperatures of the two crossing points
coincide.}}
  \end{center}
\end{figure}

The analysis of the model with competitive interactions leads to the same conclusions: the 
percolation point is $J_1/kT_p=1.1670(7)$, in accord with the 
magnetization temperature $J_1/kT_c=1.1665(9)$. In Fig. \ref{Pottsfr}
we plotted both the Binder and the percolation cumulant, so that 
we can have a visual comparison of the two thresholds.
The critical percolation indices
agree with the ones we have found for the 3-state Potts model, as we can see in
Table \ref{tab2} (Model 2).

In conclusion, we have found that the percolation point of the 
pure-site clusters of various models coincides with the 
critical point of the thermal magnetization transition. We examined theories
with discrete and continuous 
spins, involving interactions beyond the fundamental nearest-neighbour one,
and we included competitive interactions by means of antiferromagnetic
couplings. 
In particular, we stress that the effective theory of the Polyakov loop for $SU(2)$ gauge 
theory (and in general for $SU(N)$) includes many couplings,  
not only nearest-neighbour spin-spin interactions\footnote{The couplings can
involve spins which are far from each other,  
a higher number of spins (plaquette interactions, six-spins couplings,
etc.), and also self-interactions.}, and that 
such couplings are ferromagnetic but also antiferromagnetic (see \cite{okawa}).
Therefore, the fact that our result is valid even for $SU(2)$ suggests 
that it is a rather general feature of bidimensional models
with a continuous magnetization transition. It would be interesting to check
whether this property of the pure-site clusters always holds in two dimensions.
We remark as well that, even if we analyzed models with magnetization
transitions, the result is probably also true for systems undergoing
a phase transition towards antiferromagnetic ground states, since in this case 
the definition of the clusters can be trivially extended by 
joining nearest-neighbouring spins of opposite signs\footnote{In the Ising model
the equivalence of the two cases is evident.}. 

Moreover, we found that the critical percolation indices are the same
for systems in the same universality class, although they do not coincide
with the thermal indices. This shows that, at criticality, the pure-site clusters 
of all models in the same universality class are, virtually, indistinguishable
from each other as far as their size distributions are concerned (although they could
differ in their topological properties). So, in many respects, the behaviour of
the pure-site clusters of a model is as universal as other features of the model, e.g. its
critical exponents. That indicates that there is a close 
relationship between such clusters and the critical behaviour of the system.
It is indeed possible to show that the pure-site clusters of 
a model "contain" the clusters which reproduce its critical behaviour in the
sense explained at the beginning of this paper \cite{sant}.

The fact that the pure-site clusters may percolate at the critical point
also for systems different
from the Ising model was suggested in \cite{coni75}; however, this prediction 
concerned just models with ferromagnetic interactions and is a conjecture.
This work extends the result to models with competitive interactions to which
the theorems and arguments used in \cite{connap,coni75} cannot be applied. 
We guess that the property we have
illustrated here is due to the fact that the pure-site clusters of the 
Ising model, for instance, are quite compact objects. This is confirmed
\cite{CK} by the
fact that, in two dimensions, even if one randomly breaks the bonds between
nearest-neighbour
spins of the same sign with a probability $p_B=\exp(-2J/kT)$, the corresponding
site-bond clusters keep percolating at the critical temperature $T_c$. That
means that the infinite pure-site cluster at $T_c$ remains an infinite
cluster even if we randomly break many of the bonds between the spins belonging
to it, which considerably reduces its size. The
presence of antiferromagnetic couplings also acts like a bond-breaking factor,
which tends to reduce the size of the original pure-site clusters because of the
fact that they favour the presence of anti-aligned spins in the
system. Nevertheless,
if the anti-alignement interactions are weaker that the main nearest-neighbour  
ferromagnetic coupling, it is likely that they do not succeed in 
breaking the original spanning structure into finite clusters.
That could also explain the fact that the critical percolation exponents
are the same as in the original model, since we can imagine that the
antiferromagnetic interactions do not perturb enough the renormalization
group trajectories in the (percolation) coupling space, so that they would keep on 
converging to the same fixed point, exactly as it happens for the
thermal transition.
That naturally leads to the question: how small should the antiferromagnetic
couplings be compared to the fundamental ferromagnetic one, so that the
pure-site clusters still percolate at $T_c$? For the systems we have studied 
the ratio was of 1:10; we have evidence that the result holds also for a ratio
3:10, we did not present it here because the statistics is quite low.
It would be particularly interesting to check 
whether the property holds as long as the competition between the different
interactions still allows a phase transition with spin-ordering to take place.

I would like to thank my PhD advisor, professor Helmut Satz, for introducing
me in this exciting field and for his support
over the years. I also thank professor A. Coniglio for helpful
discussions. I gratefully acknowledge the financial support of
the TMR network ERBFMRX-CT-970122 and the DFG Forschergruppe FOR
339/1-2.

\begin{table}[h]
\begin{center}
\begin{tabular}{|c|c|c|c|c|c|}
\hline$\vphantom{\displaystyle\frac{1}{1}}$
&$\beta_p/\nu_p$ &$\gamma_p/\nu_p$  & $\nu_p$ & Fractal Dimension D&Cumulant at $T_p$\\
\hline
\hline$\vphantom{\displaystyle\frac{1}{1}}$
Predictions &  5/96=0.052083.. & 91/48=1.89583..&1&187/96=1.947916..&\\
\hline$\vphantom{\displaystyle\frac{1}{1}}$
Model 1 & 0.052(2) & 1.901(11)&1.004(9)&1.947(2)&0.9832(4)\\
\hline$\vphantom{\displaystyle\frac{1}{1}}$
Model 2 & 0.051(4) & 1.908(16)&1.02(4)&1.947(3) &0.9821(22)\\
\hline$\vphantom{\displaystyle\frac{1}{1}}$
Model 3 & 0.053(3) & 1.902(14)&0.99(3)&1.946(4)&0.9837(18)\\
\hline$\vphantom{\displaystyle\frac{1}{1}}$
Model 4 & 0.051(4) & 1.907(18)&0.993(35)&1.946(4)&0.9811(32)\\
\hline
\end{tabular}
\caption{\label{tab1} Critical percolation indices for the models in the 2-dimensional
  $Z(2)$ universality class.}
\end{center}
\end{table}

\begin{table}[h]
\begin{center}
\begin{tabular}{|c|c|c|c|c|c|}
\hline$\vphantom{\displaystyle\frac{1}{1}}$
&$\beta_p/\nu_p$ &$\gamma_p/\nu_p$  & $\nu_p$ & Fractal Dimension $D$ &Cumulant at $T_p$\\
\hline
\hline$\vphantom{\displaystyle\frac{1}{1}}$
Predictions & 7/80=0.0875 & 73/40=1.825 &5/6=0.8333..&153/80=1.9125 &\\
\hline$\vphantom{\displaystyle\frac{1}{1}}$
Model 1 & 0.092(11) & 1.832(18)&0.82(2)&1.910(4) &0.932(2)\\
\hline$\vphantom{\displaystyle\frac{1}{1}}$
Model 2 & 0.085(14) & 1.842(21)&0.84(3)&1.914(5)&0.929(5)\\
\hline
\end{tabular}
\caption{\label{tab2} Critical percolation indices for the models in the 2-dimensional
  $Z(3)$ universality class.}
\end{center}
\end{table}


\begin{thebibliography}{99}



\bibitem{onsager} Comment
of L. Onsager to the paper of C. J. Gorter, Nuovo Cim. Suppl. 
{\bf 6}, 249 (1949).

\bibitem{stauff} D. Stauffer, A. Aharony,
  {\it Introduction to Percolation Theory},
      Taylor {\&} Francis, London 1994.

\bibitem{connap} A. Coniglio et al., 
J. Phys. A {\bf 10}, 205 (1977).

\bibitem{sykes} M. F. Sykes, D. S. Gaunt, J. Phys. A {\bf 9},
  2131 (1976).

\bibitem{kopelman} J. Hoshen, R. Kopelman, Phys. Rev. B {\bf 14}, 3438 (1976).

\bibitem{DSM} M.\ Falcioni et al., \PL B {\bf 108}, 331 (1982); \par
E.\ Marinari, \NP B {\bf 235}, 123 (1984); \par
G.\ Bhanot et al., \PL B {\bf 183}, 331 (1986); \par
A.\ M.\ Ferrenberg, R.\ H.\ Swendsen, \PRL {\bf 61}, 2635 (1988) and {\bf 63},
1195 (1989).

\bibitem{CK} A. Coniglio, W. Klein, J. Phys. A {\bf 13}, 2775 (1980).

\bibitem{stella} A. L. Stella, C. Vanderzande, Phys. Rev. Lett. {\bf 62}, 1067 (1989).

\bibitem{santo} P. Bialas et al.,
Nuc. Phys. B {\bf 583}, 368 (2000).

\bibitem{Teper} M. Teper, Phys. Lett. B {\bf 313}, 417 (1993).

\bibitem{conpe} A. Coniglio, F. Peruggi, J. Phys. A {\bf 15}, 1873 (1982).

\bibitem{vander} C. Vanderzande, J. Phys. A {\bf 25}, L75 (1992).

\bibitem{okawa} M. Okawa, Phys. Rev. Lett. {\bf 60}, 1805 (1988).

\bibitem{sant} S. Fortunato, cond-mat/0205211.

\bibitem{coni75} A. Coniglio, J. Phys. A {\bf 8}, 1773 (1975).

\end{thebibliography}
\end{document}